# Magnetic Induction Dependence of Hall Resistance in Fractional Quantum Hall Effect

Tadashi Toyoda

*Department of Physics, Tokai University,*
*4-1-1 Kitakaname, Hiratsuka-shi, Kanagawa 259-1292 Japan*

**Abstract**

We constructed a Hall resistance formula for the fractional quantum Hall effect by analyzing the experimental data reported in [J. P. Eisenstein and H. L. Stormer, Science **248**, 1510 (1990)]. The formula is given as a function of magnetic induction, chemical potential and temperature. The Hall resistance function contains a single-electron energy spectrum, which has phenomenological perturbation terms with three tunable parameters. The formula yields 12 plateaus that are consistent with the experiment. The perturbations can be interpreted as precession or nutation of a Landau orbital in the three-dimensional space.

## 1. Introduction

Quantum Hall effects (QHE) are observed in two-dimensional electron systems realized in semiconductors [1, 2, 3, 4] and graphene [5, 6, 7, 8]. In QHE the Hall resistance exhibits plateaus as a function of magnetic induction. In the fractional quantum Hall effects (FQHE) the values of the Hall resistance on plateaus are $h/e^2$ divided by rational fractions, where $-e$ is the electron charge and $h$ is the Planck constant. The magnetic induction dependence of the Hall resistance is the strongest experimental evidence for FQHE. Nevertheless, none of existing theoretical models of FQHE can yield the Hall resistance as a function of magnetic induction that is consistent with the experiment [9]. In this work we extended the theory of the integer quantum Hall effects (IQHE) [7, 8, 10, 11] to investigate the Hall resistance in FQHE. We analyzed the experimentally measured Hall

*Email address:* `toyoda@keyaki.cc.u-tokai.ac.jp` (Tadashi Toyoda)

resistance by Eisenstein and Stormer [9], particularly the locations of fractional plateaus on the magnetic induction axis and the values of the Hall resistance on plateaus. We constructed a model for the Hall resistance as a function of magnetic induction, chemical potential and temperature. The model contains phenomenological perturbation terms in the single-electron energy spectrum. The perturbation terms successively split a Landau level into sublevels, whose reduced degeneracies cause the fractional quantization of Hall resistance. The obtained Hall resistance formula yields twelve plateaus whose locations on the magnetic induction axis are consistent with the experiment [9]. Examination of the lowest Landau level wave function in the 3-dimensional space implies the perturbation corresponds to precession or nutation of the Landau orbital in the three-dimensional space.

## 2. Hall resistance formula

Non-uniform distribution of electron density due to the Lorentz force is the essential cause of quantum Hall effects. Theoretically the non-uniform distribution can be taken into account by using the method of subsystem [7, 8, 10, 11], in which the system is theoretically divided into many strips of rectangular-shaped subsystems parallel to the direction of the bias current. The electron density in each subsystem may be different, but the chemical potential takes the same value.

To derive the Hall resistance formula we assume a model Hamiltonian of the two-dimensional electrons to be $H = H_0 + H_{\text{spin}} + H_{\text{e}} + H_{\text{int}}$, where $H_0$ is the kinetic energy term with the external perpendicular magnetic induction, $H_{\text{spin}}$ is the Zeeman spin term, $H_{\text{e}}$ is the coupling to the electric field, and $H_{\text{int}}$ is the electron-electron interaction term. Then the equations of motion for the mechanical momentum in each subsystem are [11]

$$\partial_t P_1^i = c^{-1} B \int_{\Omega^i} J_2^i - e \int_{\Omega^i} E_1^i \rho^i - \tau^{-1} e M \int_{\Omega^i} J_1^i, \qquad (1)$$

and

$$\partial_t P_2^i = -c^{-1} B \int_{\Omega^i} J_1^i - e \int_{\Omega^i} E_2^i \rho^i - \tau^{-1} e M \int_{\Omega^i} J_2^i, \qquad (2)$$

where $P_k^i$, $\rho^i$, and $J_k^i$ are the quantum statistical expectation values for the mechanical momentum, electron number density, and current density, respectively. The superscript $i$ denotes a subsystem. The integral notation is defined as $\int_{\Omega^i} =$ $\int_0^L dx_1 \int_0^{\Delta L} dx_2$, where $L$ and $\Delta L$ are the length and width of a subsystem $\Omega^i$. The electron effective mass is denoted by $M$. Magnetic induction and electric field are given as $\bm{B} = (0,0,B)$ and $\bm{E}^i = (E_1^i, E_2^i, 0)$, respectively. To ensure the Ohm's law we introduced a phenomenological damping term with a relaxation time $\tau$ [12]. We assume $\rho^i$, $J_1^i$ and $J_2^i$ in each subsystem are uniform.

To calculate the Hall resistance it is necessary to define macroscopic currents $I_k$ that correspond to experimentally measurable currents. We first define macroscopic currents $I_k^i$ in a subsystem $\Omega^i$ such that

$$\int_{\Omega^i} J_1^i = L_1 \Delta L J_1^i = L_1 I_1^i, \tag{3}$$

$$\int_{\Omega^i} J_2^i = L_1 \Delta L J_2^i = \Delta L I_2^i. \tag{4}$$

We also define the Hall voltage in each subsystem such that

$$V_2^i = -E_2^i \Delta L. \tag{5}$$

Considering the experimental conditions, we impose the steady state condition $\partial_t P_k^i = 0$, and the boundary condition $I_2^i = 0$. Then (2) and (4) give

$$I_1^i = \frac{ec}{B} V_2^i \rho^i, \tag{6}$$

which holds for each subsystem. By adding (6) from all subsystems, we obtain

$$I_1 = \frac{ec}{B} \sum_i V_2^i \rho^i, \tag{7}$$

where $I_1 = \sum_i I_1^i$ is the experimentally measurable macroscopic current.

We assume the expectation value for the electron number density is given in terms of the Fermi distribution function $f(\varepsilon_q + \delta\varepsilon^i; T) = [1 + \exp\{(\varepsilon_q + \delta\varepsilon^i - \mu)/k_B T\}]^{-1}$, where $k_B$, $T$, and $\mu$ are the Boltzmann constant, temperature, and chemical potential, respectively. The electron energy spectrum in the subsystem $\Omega^i$ consists of an $i$-independent part $\varepsilon_q$ and an $i$-dependent part $\delta\varepsilon_i$, where $q$ is the quantum number of a quasi-electron state. The total current $I_1$ is

$$I_1 = ecB^{-1} \sum_i V_2^i \sum_q D(q) f(\varepsilon_q + \delta\varepsilon^i; T), \tag{8}$$

where $D(q)$ is the degeneracy of the energy level $q$. The experimentally measurable Hall potential difference is $V_2 = \sum_i V_2^i$. In general the presence of $\delta\varepsilon^i$ in the Fermi distribution prohibits the evaluation of the sum over $i$ to obtain $V_2/I_1$. However, if $\delta\varepsilon^i$ is much smaller than the smallest increment of energy level $\varepsilon_q$, then the summation of the Fermi distributions is possible. After the summation over $i$ we find

$$I_1 = ecB^{-1} V_2 \sum_q D(q) f(\varepsilon_q; T) \ . \tag{9}$$

This yields the inverse of Hall resistance

$$R_\mathrm{H}^{-1} = ecB^{-1} \sum_q D(q) \left\{1 + \exp\left[(\varepsilon_q - \mu)/k_B T\right]\right\}^{-1} . \tag{10}$$

The single-electron energy spectrum is given by $\varepsilon_q$ with a quantum number $q$. The degeneracy of energy level $q$ is denoted by $D(q)$.

### 3. Model of FQHE

To construct a model for the FQHE let us first examine the theoretical mechanism of plateaus in IQHE, which can be quantitatively explained by adopting the Landau level $\varepsilon_q = \varepsilon_{N\alpha} = \hbar\omega_c(N + 1/2 + \zeta\alpha)$ as the energy spectrum in (10) [7, 8, 10, 11]. Here $\omega_c = eB/Mc$ is the cyclotron frequency. The quantum number $N$ is a non-negative integer. The spin variable $\alpha$ takes the values $\pm 1$. The Zeeman spin term is $\hbar\omega_c \zeta\alpha$ with $\zeta = (g^*/2)(M/2M_0)$, where $M_0$ is the electron rest mass. The degeneracy of a Landau level with a given spin variable is

$$D(N, \alpha) = eB/hc \equiv D_0. \tag{11}$$

The magnetic induction $B$ in this $D_0$ cancels the $B$-dependence of the factor $ecB^{-1}$ in (10). Consequently, the inverse of Hall resistance for IQHE becomes

$$R_\mathrm{H}^{-1} = e^2 h^{-1} \sum_{N=0}^{\infty} \sum_\alpha \left\{1 + \exp\left[(\varepsilon_{N\alpha} - \mu)/k_B T\right]\right\}^{-1} . \tag{12}$$

Because of the Fermi distribution the inverse of Hall resistance becomes a sum of step functions in the zero temperature limit,

$$\lim_{T \to 0} R_\mathrm{H}^{-1} = e^2/h \sum_{N=0}^{\infty} \sum_\alpha \theta\left(B_{N\alpha} - B\right). \tag{13}$$

The locations of step edges on the $B$ axis are given by

$$B_{N\alpha} = (\mu Mc/e\hbar)\{N + 1/2 + \alpha\zeta\}^{-1}. \tag{14}$$

Calculation of (13) from (10) shows the quantization unit of $R_{\mathrm{H}}^{-1}$ is $e^2/h$ because of the degeneracy of a Landau level $D_0$. That is,

$$ecB^{-1}D_0 = ecB^{-1}(eB/hc) = e^2/h. \tag{15}$$

Let us inspect the Hall resistance data in FQHE experiment. The quantization unit of $R_{\mathrm{H}}^{-1}$ on $e^2/3h$, $2e^2/3h$ and $4e^2/3h$ plateaus observed in the FQHE experiment [9] is $e^2/3h$. In view of (15) the most plausible explanation for this is that a Landau level is split into three sublevels. Each sublevel has the degeneracy $D_1 = D_0/3$. We assume that the level-splitting is caused by a perturbation Hamiltonian $\mathcal{H}'_1$, which yields the new quantum numbers $m_1 = -1, 0, 1$ for sublevels. Let us call these sublevels the $m_1$ sublevels.

The $2e^2/5h$, $3e^2/5h$, $4e^2/5h$, and $7e^2/5h$ plateaus in FQHE can be explained by assuming an additional perturbation Hamiltonian $\mathcal{H}'_2$ that splits each $m_1$ sublevel into five sublevels. Let us call these sublevels the $m_2$ sublevels. Each sublevel has the degeneracy $D_2 = D_1/5$. We assume that $\mathcal{H}'_2$ is small perturbation to $\mathcal{H}'_1$.

The $3e^2/7h$ and $4e^2/7h$ plateaus in FQHE can be explained by assuming an additional perturbation Hamiltonian $\mathcal{H}'_3$ that splits each $m_2$ sublevel into seven sublevels. Let us call these sublevels the $m_3$ sublevels. Each sublevel has the degeneracy $D_3 = D_2/7$. We assume that $\mathcal{H}'_3$ is small perturbation to $\mathcal{H}'_2$.

Hence, the quantized values of FQHE resistance at fractional plateaus can be attributed to the degeneracies of sequentially split sublevels. This analysis indicates a model energy spectrum

$$\varepsilon(N, \alpha, m) = \varepsilon_{N\alpha} + \hbar\omega_c\left\{\lambda_1 m_1 + \lambda_2 m_2 + \lambda_3 m_3\right\}, \tag{16}$$

where $m_l$ is an integer ranging $-l \leq m_l \leq l$. The parameters $\lambda_l$ are assumed to be $|\lambda_{l+1}| < |\lambda_l|$. We have defined $m = (m_1, m_2, m_3)$.

Using the Hall resistance formula (10), we can determine the parameters $\lambda_l$ from the experiment. In the zero-temperature limit, the locations of step edges on the $B$ axis are given by (14) as

$$\begin{aligned}B_{N\alpha m}^{\mathrm{FQHE}} =& (\mu Mc/e\hbar) \\ & \times \{N + 1/2 + \alpha\zeta + \lambda_1 m_1 + \lambda_2 m_2 + \lambda_3 m_3\}^{-1}.\end{aligned} \tag{17}$$

By reading the values of $B_{N\alpha}$ from the experimental Hall resistance data at very low temperatures, it is possible to determine $\lambda_l$.

Because the number of possible $m_l$'s for a given $l$ is $2l+1$, the degeneracy of an energy level with quantum numbers $(N, \alpha, m)$ is $D(N, \alpha, m) = D_0(N, \alpha) \prod_{l=1}^{3} (2l+1)^{-1}$. Hence the inverse of Hall resistance for FQHE is given as

$$R_{\text{H}}^{-1} = e^2 h^{-1} \sum_{N=0}^{\infty} \sum_{\alpha} \prod_{l=1}^{3} (2l+1)^{-1}$$
$$\times \sum_{m} \left\{ 1 + \exp\left[ (\varepsilon(N, \alpha, m) - \mu)/k_B T \right] \right\}^{-1}, \tag{18}$$

where we have defined $\sum_m = \sum_{m_1} \sum_{m_2} \sum_{m_3}$. This formula yields the values of Hall resistance on plateaus as

$$R_{\text{H}} = \frac{h}{e^2} \frac{j}{3 \cdot 5 \cdot 7} \qquad (j = 1, 2, ...). \tag{19}$$

The Hall resistance given by (18) is plotted as a function of $B$ in Fig. 1. The three parameters $\lambda_l$ in (16) are fitted to the experimental Hall resistance curve in Ref. [9]. Their values are $\lambda_1 = 0.25$, $\lambda_2 = 0.14$, and $\lambda_3 = 0.003$. Considering the Hall resistance data for the IQHE experiment in Ref. [13], the effective g-factor is adjusted to $g^* = 12$. The effective mass is $M = 0.067 M_0$. The chemical potential is determined by the slope of experimental Hall resistance curve for weak magnetic induction. The value is $\mu = 13.14 \times 10^{-15}$ erg. The theoretical resistance curve in Fig. 1 is calculated for $T = 85$ mK which is the experimental temperature in Ref. [9].

In order to see the plateaus clearly the theoretical resistance curve for $T = 5$ mK is plotted in Fig. 2. The experimentally observed quantized Hall resistance plateaus $1/3$, $2/5$, $3/7$, $4/7$, $3/5$, $2/3$, $4/5$, $1$, $4/3$, $7/5$, $5/3$, and $2$ perfectly agree with theoretical results, and are indicated by arrows in Fig. 2. Although the experimentally observed plateaus $4/9$, $5/9$ and $7/9$ are not exactly produced theoretically, the theory yields the corresponding plateaus $47/105$, $58/105$, and $82/105$. The difference between the Hall resistances associated to the experimental three plateaus and theoretical plateaus are less than a few percent.

In Fig. 3 the magnetic induction and temperature dependence of the Hall resistance is shown in a 3D plot. It shows the Hall resistance curve given by the formula (18) becomes classical as temperature increases. Hence the formula (18) can yield IQHE, FQHE and classical Hall effects.

## 4. Angular momentum of lowest Landau level wave function

The quantum number $m_l$ introduced in the model perturbation energy spectrum (16) ranges $-l \leq m_l \leq l$. Therefore, it is plausible that these quantum numbers $m_l$ and $l$ correspond to angular momentum. Because the orbital angular momentum operator cannot be defined in the 2-dimensional space, on which the lowest energy Landau level wave function $\Phi_0(\rho, \phi)$ is calculated, it is necessary to consider the problem in the 3-dimensional space. The spatial dimension can be extended by changing the two-dimensional polar coordinates $(\rho, \phi)$ to the three-dimensional polar coordinates $(r, \theta, \phi)$. It is also necessary to consider explicitly the confining potential wave function $\chi_0$. We assume $\chi_0(r\cos\theta) = (\sqrt{2\pi}d)^{-1}\exp(-r^2\cos^2\theta/2d^2)$, where $d$ is the thickness of the 2-dimensional system. By adopting the vector potential $\boldsymbol{A} = (-Bx_2/2, Bx_1/2, 0)$, the 3-dimensional lowest Landau level wave function can be written as

$$\Phi_0 = N_m \left(\frac{d^2}{a^2}\right)^{|m|/2} \exp\left[-z^2\right] \sum_{j=0} \frac{1}{j!}\left(1 - \frac{d^2}{2a^2}\right)^j$$

$$\times \frac{(z^2)^{(|m|+2j)/2}}{(2(|m|+2j)-1)!!} \sum_{l=0}^{\infty} C(l,m;j) Y_{lm}(\theta, \phi) \qquad (20)$$

where $N_m$ is the normalization factor, $a = \sqrt{c\hbar/eB}$ is the magnetic length, $Y_{lm}$ is spherical harmonics and $z^2 = r^2/2d^2$. The expansion coefficient $C(l, m; j)$ is given by

$$C(l, m; j) = \sqrt{2\pi}\left[\frac{2l+1}{2}\frac{(l-|m|)!}{(l+|m|)!}\right]^{1/2}$$

$$\times \int_{-1}^{1} dx P_l^{|m|}(x) P_{|m|+2j}^{|m|+2j}(x) . \qquad (21)$$

This expansion shows that the lowest Landau level in the three-dimensional space is a superposition of angular momentum eigenstates of different $l$. The allowed values of $m$ in (20) are only non-positive integers [14]. Because the quantum number $m_l$ ranges from $l$ to $-l$, it cannot belong to the unperturbed state given by (20). Therefore, the quantum number $m_l$ may correspond to new rotational degree of freedom resulted from precession or nutation of the Landau orbital.

## 5. Concluding remarks

We explained the fractional quantized values of the Hall resistance on plateaus in terms of the degeneracies of sublevels created from Landau levels by the phe-nomenologically introduced perturbation terms in the single-electron energy spectrum. The angular momentum nature of perturbation implies precession or nutation phenomena.

The obtained Hall resistance formula yields twelve plateaus whose locations on the $R_\mathrm{H} - B$ plot are consistent with the experiment. No existing theories can yield this quantitative fit to the experiment. In this model only three tunable parameters were adjusted. We also succeeded to show the temperature dependence of the Hall resistance. The 3D-plot graph shows how FQHE disappears and becomes classical Hall effect with explicit temperature dependence. The Hall resistance formula (10) is valid for IQHE and FQHE. The formula shows the Hall resistance depends only on the single-electron energy spectrum via Fermi distribution. This indicates that the Fermi liquid theory [15, 16] is valid for IQHE and FQHE.

**Acknowledgements**
I thank M. Yasue for thorough discussion. I thank M. Fujita, K. Yamada, and T. Uchida for comments.

**Figure Captions**

Fig. 1
Theoretical Hall resistance as a function of magnetic induction $B$ at $T = 85$ mK calculated by the formula (18) is plotted in blue. The experimental Hall resistance [9] at $T = 85$ mK is also plotted in gray.

Fig. 2
Theoretical Hall resistance as a function of $B$ at $T = 5$ mK calculated by the formula (18) is plotted in blue. Experimental Hall resistance [9] at $T = 85$ mK is also shown in gray. The horizontal arrows indicate plateaus.

Fig. 3
Theoretical Hall resistance as a function of $B$ and $T$ calculated by the formula (18) is shown as a 3D plot for $0 < T < 10$ K and $0 < B < 30$ T.

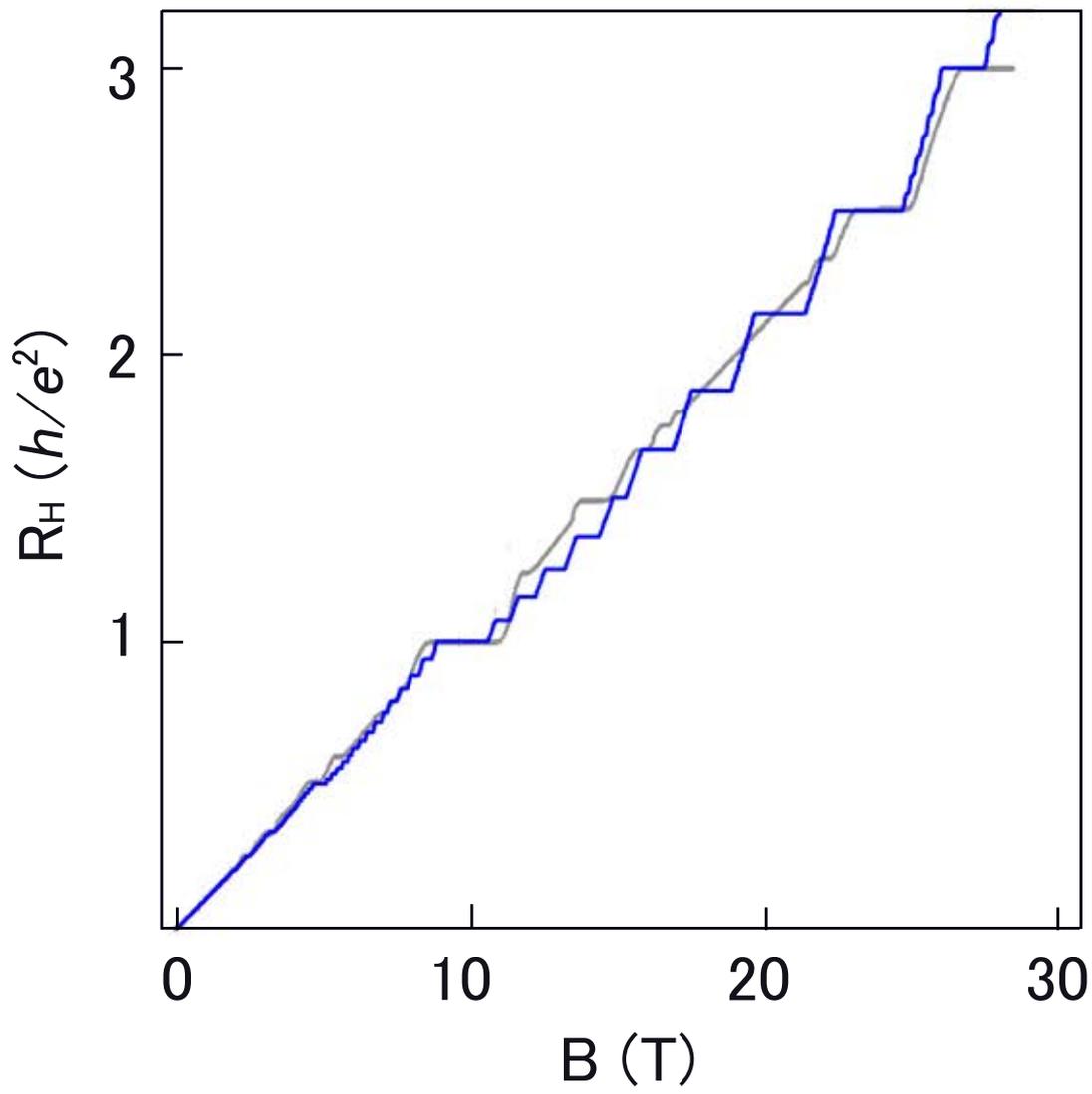

Fig. 1

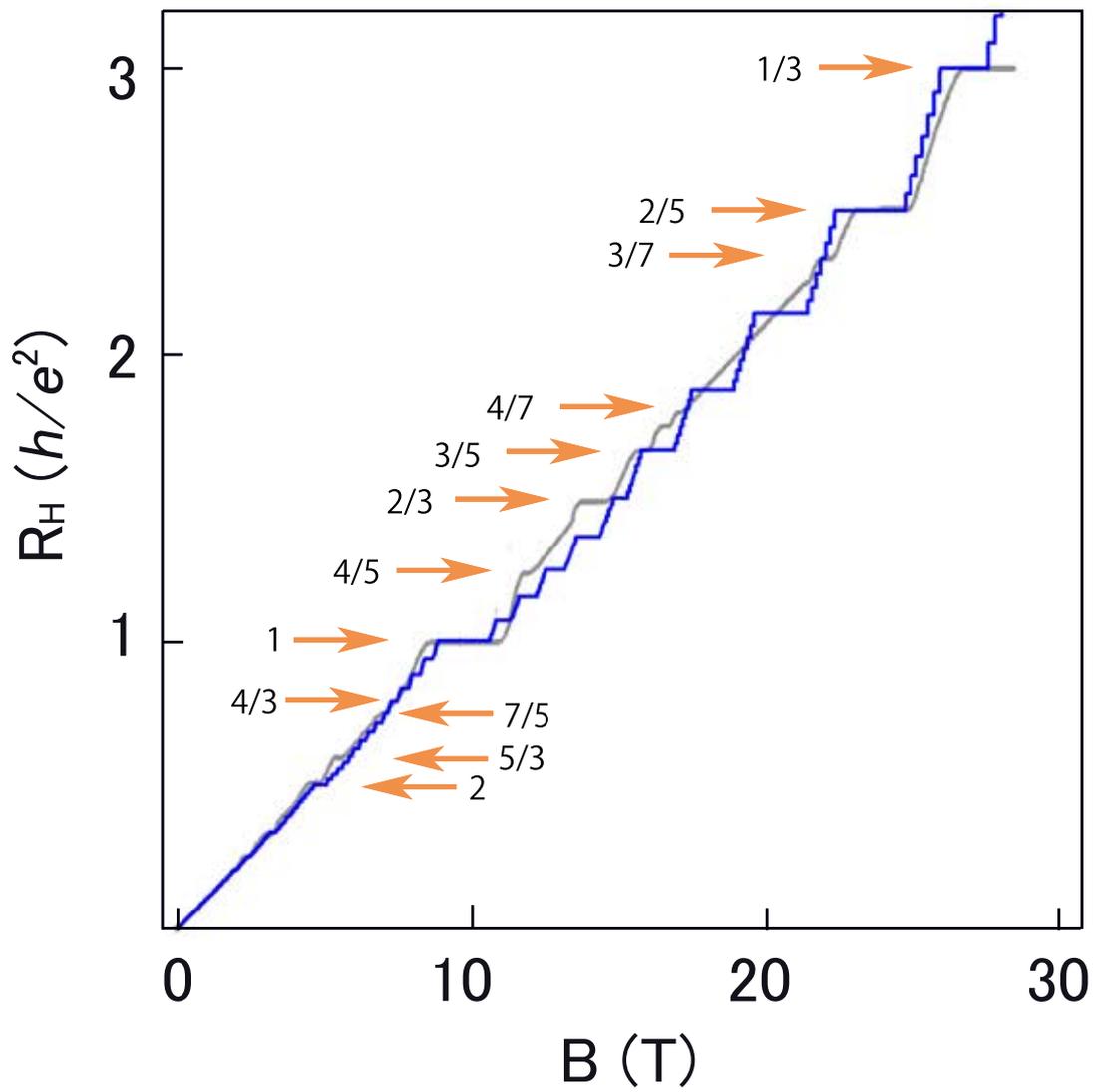

Fig. 2

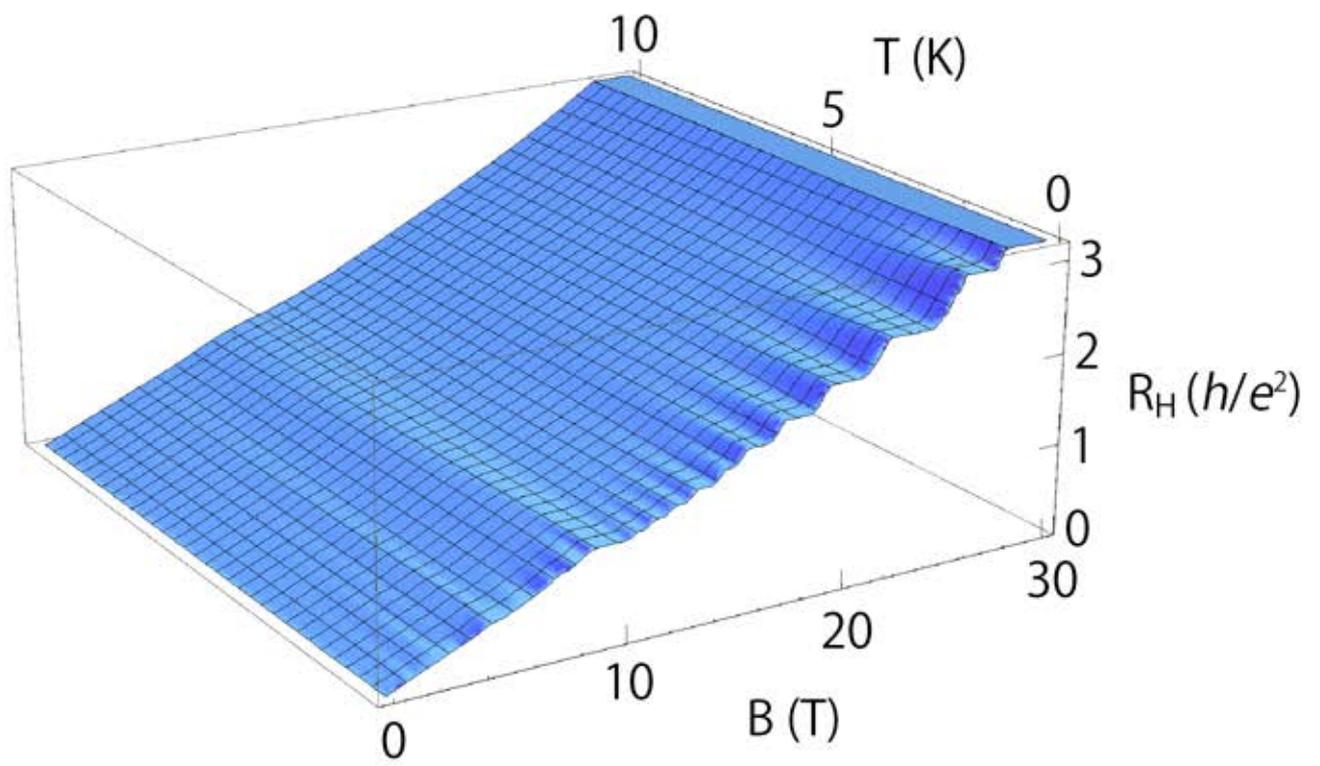

Fig. 3